\documentclass[preprint2]{aastex}
\usepackage{natbib,latexsym}
\usepackage{lscape}
\usepackage{rotating}
\usepackage{multirow}
\usepackage{verbatim}

\begin{document}

\title{HI Observations of the Ca~II absorbing galaxies Mrk~1456 and SDSS~J211701.26-002633.7}

\author{B. Cherinka\altaffilmark{1},
  R.E. Schulte-Ladbeck\altaffilmark{1}, and J. L. Rosenberg\altaffilmark{2}}
\affil{Department of Physics \& Astronomy, University of Pittsburgh,
    Pittsburgh,PA 15260}
\affil{Department of Physics and Astronomy, George Mason University,
  Fairfax, VA 22030}

\email{bac29@pitt.edu}

\begin{abstract}
In an effort to study Damped Lyman Alpha galaxies at low
redshift, we have been using the Sloan Digital Sky Survey to identify galaxies
projected onto QSO sightlines and to characterize their optical
properties.  For low redshift galaxies, the HI~21cm emission line can be
used as an alternate tool for identifying possible DLA galaxies, since
HI emitting galaxies typically exhibit HI columns that are larger than
the classical DLA limit.  Here we report on follow-up HI 21~cm emission line observations of two
DLA candidates that are both low-redshift
spiral galaxies, Mrk~1456 and SDSS~J211701.26-002633.7. The
observations were made using the Green Bank and Arecibo Telescopes, respectively.  Analysis of their HI properties
reveal the galaxies to be about one and two $M_{HI}^*$ galaxies, respectively, and to have average HI mass,
gas-richness, and gas mass fraction for their morphological types. We consider
Mrk~1456 and SDSS~J211701.26-002633.7 to be candidate DLA systems based upon
the strength of the CaII absorption lines they cause in their QSO's spectra, and impact
parameters to the QSO that are smaller than the stellar disk.
Compared to the small numbers of other HI-detected DLA and candidate DLA galaxies, Mrk~1456 and
SDSS~J211701.26-002633.7 have high HI masses.  Mrk~1456 and SDSS~J211701.26-002633.7 have also been
found to lie in galaxy groups that are high in HI gas mass compared to the group containing SBS~1543+593, the only
DLA galaxy previously known to be situated in a galaxy group.  When compared with
the expected properties of low-z DLAs from an HI-detected sample of galaxies, Mrk~1456 and
SDSS~J211701.26-002633.7 fall within the ranges for impact
parameter and $M_B$; and the HI mass distribution for the HI-detected
DLAs agrees with that of the expected HI mass distribution for low-z
DLAs.  Our observations support galaxy-evolution models
in which high mass galaxies make up an increasing contribution to the DLA cross-section at lower redshifts.  
We also report on the 21~cm line emission of Mrk~1457, a
Seyfert galaxy observed within the beam of the Green Bank Telescope.
\end{abstract}





\section{Introduction}
Damped Lyman Alpha (DLA) absorbers are the primary reservoir of cool,
neutral gas in the Universe and the ability to detect them easily over
a wide range of redshifts make them an important tracer of the
evolution of gas in galaxies or protogalaxies over cosmic time.  DLA
systems are identified at redshifts 0$<$z$<$6 in the spectra of QSOs \citep{wolf}.
Identified from the strong n=1 to n=2 absorption line of HI, the
detection of these absorption-selected objects depends primarily
on the brightness of the background QSO rather than on emission from
stars or gas. It is noteworthy that the column density of
$N_{\rm HI} > 10^{20.3}$ atoms~cm$^{-2}$, that defines a
DLA \citep{wolf2}, is similar to the threshold column
density required for the onset of star formation in the
local Universe \citep{kenn98}.  DLAs, then, should be associated with
present-day visible galaxies \citep{wolf2,proch97,bois03}.  However,
difficulty still remains in connecting observations of high-redshift
DLAs with those of low-redshift DLAs, as well as connecting
galaxy-evolution simulations with DLA observations.  Simulations at
low redshift by \citet{okoshi} and \citet{naga} indicate that the DLA incidence is dominated by more compact, faint
galaxies with a narrow impact parameter distribution ($<$3 kpc),
rather than by large disk galaxies.  Observationally, low redshift DLA systems have been shown to consist  
of galaxies spanning a wide range of luminosities, surface  
brightnesses, impact parameters, and morphologies \citep{lebrun97,chen03,rao03}.  At higher redshifts, the associated
host galaxy to the DLA has proven difficult to identify due to the
high brightness contrast between the quasar and the absorbing galaxy.  Another issue in the identification of the host galaxy at high redshift is the fact that one can no longer detect faint galaxies that may be intercepting your sightline.  This may cause a misidentification of the host galaxy, especially in high density regions like groups. This may contribute to
the larger impact parameters being measured for high redshift
systems.  \citet{christ} identified candidate hosts to high-redshift (2$<$z$<$4)
DLAs through IFU observations of the Ly$\alpha$-emission line. While
they have a small sample size, they find an average impact
parameter of 16 kpc, and suggest the distribution of HI clouds for DLAs
extends far beyond the optical sizes of dwarfs.  Simulations by
\citet{gardner} resulted in larger impact parameters (10-15 kpc) for
2$<$z$<$4.
   
Selecting systems by identifying damped Lyman-$\alpha$ absorption provides only a pencil-beam examination of the gas content of DLA host galaxies, missing the full gas content that can be probed.  An
alternative then, to searching for gas through Lyman-alpha absorption, is to use the 21~cm emission line, which probes the global gas content in galaxies, to search for systems that are likely to show damped Lyman-$\alpha$ profiles if there were a background source.  The HI hyperfine structure line at 21~cm is very weak,
and its detection is currently limited to low redshifts.  HI 21~cm observations probe the cold, neutral Hydrogen gas in a galaxy and come in two forms, 21~cm absorption and 21~cm emission.  21~cm absorption, like other absorption-line probes, provides a pencil-beam look at the HI content of a galaxy.  It requires a background radio-loud point source and allows for a direct determination of $N_{HI}$.  21~cm line emission is a probe of the gas contained within the beam of the telescope, smoothed according to the spatial resolution of the telescope.  The resulting HI emission spectrum reveals the global HI content of the galaxy and allows for determination of the HI mass while maps show the distribution and cross-sections of HI gas around a galaxy.  Estimates of the gas column density within the beam can be obtained by the ratio of the HI flux to the area covered.  A probe of emission might reveal an average HI column above $10^{20.3}$, however this is over a much larger area than that probed by absorption, therefore there is no guarantee that a QSO probe will reveal a DLA at every position within the emission beam.  Emission-line probes above $10^{20.3}$ only say that somewhere they generally exceed the DLA limit.  \citet{ryan03} showed that HI-emission selected galaxies generally show more cross-section at higher column densities than they do at lower column densities and most exhibit column densities that are above the DLA limit.  The HI-emission-selected galaxies are consistent with the hypothesis that the local galaxy population can explain the properties of the
local DLAs.  \citet{rosen03}, and \citet{ryan03} studied the expected properties of the z=0 DLA population using
blind 21~cm emission surveys, and found that, while a large fraction of HI-selected galaxies are dwarf or low surface brightness galaxies, there is no need for an optically invisible population of galaxies to explain the HI population or to contribute significantly to the DLA cross section.    

There are presently only a small number of (sub) DLAs or DLA candidates known at
low-redshift and an even smaller number that have HI measured in
emission.   It should be noted that a DLA (and sub-DLA) system is classically defined as a system with an observed $N_{\rm HI} > 10^{20.3}$~atoms~cm$^{-2}$ ($10^{19.0}$~cm$^{-2}$ $<$ $N_{HI}$ $<$ $10^{20.3}$~cm$^{-2}$) measured via the Ly-$\alpha$ absorption line.  The term ``candidate DLA'' is used to describe all galaxies that have $N_{\rm HI} > 10^{20.3}$ atoms~cm$^{-2}$ measured through other means, or those with no $N_{HI}$ measurement but meet the criteria of various metal-DLA relationships (e.g. Mg~II,Ca~II,Na~I).  \citet{zwaan} compared the properties of 20 low-redshift
(z$<$1) DLA/sub-DLA galaxies with a sample of local, optically
selected galaxies studied in HI 21~cm.  Only
2 of the 20 DLAs, SBS~1543+593 and NGC~4203, have measured HI 21~cm emission.  One other DLA, the absorber towards the QSO OI~363, and two sub-DLAs, the absorbers towards the QSOs PKS~0439-433 and PG~1216+069, have
been searched for 21~cm emission.  We use the HI observations of these objects, along with the ones reported here, to
further investigate the properties of DLAs at z=0 and expand on the
number of DLAs or candidates known at low-z with HI 21~cm emission
measurements.

Table 1 lists the HI and optical properties of the
known/candidate DLAs.  A summary of the HI 21~cm observations of these five known/candidate DLAs is as follows.  The z=0.009 dwarf galaxy SBS~1543+593 has an HI column density of log $N_{HI}$=20.41 \citep{bowen05} as measured by the Lyman~$\alpha$ line, while \citet{cheng} find a column density of log $N_{HI}=20.69$
from the 21~cm emission mapped at the position of the QSO.  
The z=0.004 S0 galaxy NGC~4203 was first mapped at 21~cm before the presence
of the background QSO Ton~1480 was revealed by X-ray observations.
The HI absorption in the X-ray spectrum of Ton~1480 suggests a
sufficiently high column density (log $N_{HI}=20.34$) to make this
system a candidate DLA.  Another known DLA, the z=0.0912 absorber towards B~0738+313 (OI~363), has been
searched for 21~cm emission but was undetected \citep{lane00}.  Two sub-DLA's have also been searched for 21~cm emission.  The z=0.101
absorber towards PKS~0439-433 was thought to be a candidate DLA
system based on its X-ray spectrum but measurement of the
Lyman~$\alpha$ line \citep{chen05} revealed a column
density (log $N_{HI}=19.85\pm0.04$) just below the limit for DLAs and it was
undetected in 21~cm emission \citep{kane01}.  \citet{briggs06}
report a weak 21~cm emission detection along the line of
sight towards the QSO PG~1216+069, which is consistent with the velocity of the
z=0.0063 sub-DLA absorber.  

In this paper we present the 21~cm emission detection of Mrk~1456 with the Green
Bank Telescope (GBT) and of SDSS~J211701.26-002633.7 with the
Arecibo telescope\footnote[1]{The Arecibo Observatory is part of the
  National Astronomy and Ionosphere Center, which is operated by
  Cornell University under a cooperative agreement with the National
  Science Foundation.}.  Section 2 describes the sample selection,
observations, data reductions and analysis.  In section 3, we discuss
the HI properties of our sample and compare them to HI-detected DLAs
and HI-selected galaxies.  We adopt a
cosmology of $H_o$=71~km~$s^{-1} Mpc^{-1}$, $\Omega_m$=0.27, $\Omega_{\Lambda}$=0.73 throughout
this paper.  

\section{Observations and Data Analysis}

\subsection{Sample Selection and Optical Data}

A search was made in the Sloan Digital Sky Survey (SDSS) for
candidate DLAs by cross-referencing \it{known }\normalfont spectroscopic galaxies
with the SDSS Data Release 5 (DR5) QSO catalog \citet{schneider07}, creating a
subset of galaxy-QSO pairs.  The motivation behind this approach was
spawned by the difficulty of identifying DLA host galaxies at both high and
low redshifts.  The galaxy-QSO pairs were further trimmed by
requiring the impact parameters to be less than twice the Petrosian
radius of the galaxy, and that $z_{QSO}$ $>$ $z_{gal}$.  Twice the
Petrosian radius was chosen as HI disks have typically been shown to
extend anywhere from 1.5 - 2.0 times the optical radius for spiral
galaxies \citep{cayatte94,broeils97} and we adopted this cutoff to
ensure inclusion of a significant amount of HI in the matching
process.  As the SDSS is a magnitude limited survey \citep{strauss02}, our selection method favors giant over dwarf
DLA candidates and thus may tend to select high HI mass galaxies.
Mrk~1456 and SDSS~J211701.26-002633.7 (SDSS~21-00 for brevity) were selected
from this set because the background QSOs are within the stellar disks
of the galaxy.  Mrk~1456 has an impact parameter, b=4.9 kpc, compared with the galaxy's r-band
Petrosian radius, r=7.1 kpc.  For SDSS~21-00, b=5.7 kpc while r=7.5
kpc.  These values are both less than the median impact parameters for
DLA hosts (b$<$8.0 kpc) as found by \citet{zwaan}.  The SDSS redshifts
reported for Mrk~1456, Mrk~1457, and SDSS~21-00 throughout the paper are
optical emission line redshifts.  

Mrk~1456 and SDSS~21-00 were chosen for follow-up HI
observations because they are strong DLA candidates that are within
the redshift limits of HI 21~cm observations.  The beam of our HI
observations also included Mrk~1457 so we present the data for this
galaxy as well.  

Direct measurement of the damped Lyman-$\alpha$ line in low-z systems
requires UV spectroscopy from space, which has not been carried out on
the two QSOs in this study.  However, the small QSO impact parameters, and the strength of the Ca~II
doublets ($\lambda\lambda$3934,3969) in both Mrk~1456 and SDSS~21-00
indicate that they are likely to be DLAs.  \citet{wild05} and \citet{wild06} have estimated
that systems with a rest EW($\lambda$3934)$\geq$0.5$\AA$ have a number
density $\sim$20-30$\%$ of DLAs and those with
EW$\geq$0.68$\AA$ are highly likely to be a subset of
the DLA population.  \citet{nestor08} note that while strong Ca~II absorbers
are likely to be DLAs, not all DLAs will have strong Ca~II EWs.
\citet{koenig} reported a Ca~II~K EW of 1.24$\pm$0.15$\AA$ for Mrk~1456, well above the threshold for
DLA candidates.  For SDSS~21-00, we measure a rest Ca~II~K EW
of 1.1$\pm$0.2$\AA$, also above threshold.  The Ca~II EW measurements
for both Mrk~1456 and SDSS~21-00 were made by fitting Gaussians to the
lines using a local fit to the continuum. 

\citet{koenig} reported on Mrk~1456 as a typical,
$L_*$, giant spiral galaxy with a spectrum of an Sc type disk galaxy.
For SDSS~21-00, comparison of it's spectrum with the templates of \citet{kenn} and using the inverse
concentration index and u-r color \citep{park05,shima01} suggests that
it is an Sb type spiral galaxy.  A morphological examination of the
optical image for SDSS~21-00 shows it is a late-type disk galaxy.  Mrk~1457 is
classified as a Seyfert 2 galaxy in NED\footnote[2]{NASA Extragalactic
  Database} and lies 2.3'(~2.2 kpc) south of Mrk~1456.  

The inclinations, i, of
each galaxy were computed using the minor-to-major axis ratio in the r band
measured from an exponential fit to the profile, measured out to a
radius of three times the effective radius.  This gives 45$^{\circ}$, 25$^{\circ}$, 
and 52$^{\circ}$, for the inclinations of Mrk~1456, Mrk~1457, 
and SDSS~21-00 respectively.

We use the SDSS g and r photometry to calculate the galaxies' absolute
magnitudes and luminosities and then transform to
Johnson-Morgan-Cousins B using the \citet{smith} transformation laws.
The absolute B-band magnitudes reported here were converted to the AB zeropoint system, corrected for galactic
extinction \citep{schlegel}, and k-corrected.  In
order to make a consistent comparison of  Mrk~1456 and SDSS~21-00 to
objects published in the literature, we use k-corrections from
\citet{pog}.  With the galaxy inclinations given above, and the HI velocity widths
listed in Table 3, we correct for internal extinction using the method of \citet{tully98}.
All magnitudes and luminosities are listed in Table 2.  Assuming $M_B^*$=-20.9
\citep{mar99}, Mrk~1456 and SDSS~21-00 are $\sim$0.5$L_B^*$
galaxies, and Mrk~1457 is a 1.4$L_B^*$ galaxy.  

Following the prescription of \citet{hopkins}, we derive star-formation rates (SFR) for our
objects and for those literature objects for which the data were
available or could be calculated (see Table 2).  The star-formation rates for Mrk~1456, and SDSS~21-00 are
global Petrosian u-band SFRs corrected for internal extinction using the Balmer-line ratio
and a Calzetti obscuration curve \citep{calzetti}.  For Mrk~1456 we adopt the
Balmer-line ratio from \citet{koenig}.   For SDSS~21-00, we
measure a Balmer-line ratio of 7.2 using the H$\alpha$ flux, corrected for
absorption via \citet{hopkins}, and the H$\beta$ flux, with the absorption and
emission components of the line deblended with two Gaussians.  The u-band absolute
magnitudes and star-formation rates are given in Table 2.  We derive
3.1$\pm$0.2 and 5.6$\pm$1.2 $M_{\odot}yr^{-1}$ for Mrk~1456 and
SDSS~21-00, respectively, both in good agreement with the average
SFRs for late-type spirals \citep{kenn83,kenn98}.  

Table 2 lists several emission-line measured, and derived, properties for
Mrk~1456 and SDSS~21-00.  For Mrk~1456, we adopt the values from
\citet{koenig} for emission-line fluxes and derived abundances.  For SDSS~21-00, emission-line fluxes were measured by
fitting Gaussians to the OIII, $H\beta$, NII, and $H\alpha$
emission-lines, while oxygen abundances were measured using the
strong-line indices of $\frac{OIII[\lambda
    5007]/H\beta}{NII[\lambda 6583]/H\beta}$ (O3N2), $\frac{NII[\lambda
    6583]}{H\alpha}$ (N2) \citep{pettini} and R23 \citep{kobul,mcgaugh}.
We find that the O3N2 and N2 derived abundances are in good agreement
with each other.  Assuming a solar abundance of 8.74$\pm$0.08 from \citet{holweger}, SDSS~21-00
has a metallicity near the solar value.  



\subsection{Green Bank Telescope Observations and Reduction}

The observations of Mrk~1456 were made on 2006 August 27 with the Green Bank Telescope, a 100-m
diameter single dish telescope with a FWHM beamwidth of
12.4$^{\prime}$/f(GHz).  The L-band Gregorian receiver was used, which
covers a frequency range of 1.15-1.73 GHz, giving a FWHM beamwidth of     
9$^{\prime}$ with two linear polarizations.  The gain on the L-band receiver is 2 K/Jy.  The
backend used was the GBT Spectrometer in the narrow bandwidth, high
resolution mode.  The spectrometer was set for a 12.5 MHz bandwidth,
with 1 spectral window.  9-level sampling was employed
to increase sensitivity and to improve the ability
to excise Radio Frequency Interference (RFI).  The raw spectral
resolution was 0.38146 kHz over 32768 channels.  We obtained 42
on/off pairs each with a 5 minute integration time resulting in a
total integration time of 7 hours, including both signal and
reference observations.     

To reduce the data we (1) removed any Radio Frequency Interference
(RFI) identified by eye in individual scans,
(2) subtracted off a baseline, and (3) averaged the two
polarizations to achieve the final spectrum.  RFI usually
presents itself as sharp spikes in the data.  Each of the
84 scans was investigated, by eye, for any such spikes and, if found,
the bad data were removed.  During the RFI removal, two other problems were discovered in the
data, including ``Bad Lag''.  ``Bad Lag'' refers to the incorrect
scaling of the raw lags by the spectrometer (O'Neil, K. 2006)\footnote[3]{http://wiki.gb.nrao.edu/bin/view/Software/ModificationRequest15C306}.  Each record,
2.5 seconds of data, where this ``Bad Lag'' occured were removed from
the scan.  In addition, when the power spectrum of the data was
analyzed, we found a sharp peak at a frequency corresponding to a
standing wave on the antenna.  The standing wave
arises from light that is reflected back and forth between the
reflector and the panel gaps on the telescope \citep{fish03}.
Due to these standing waves, fourteen scans (140 minutes), were
removed from the 1st polarization and eight scans (80 minutes)
were removed from the 2nd polarization.  
Once the bad data were removed, a 3rd, 5th, or 7th order polynomial was fit to each scan and
subtracted.  After the subtraction of the baseline, each scan was then
Hanning smoothed  and then further smoothed with a 500 channel (45 km $s^{-1}$)
boxcar filter.   After smoothing the data, a weighted average of all
of the scans in both polarizations was computed.  Each scan is
assigned a weight based on it's antenna temperature, exposure time,
and frequency resolution.  Since an entire scan gets a
single weight, the channels in a scan that were flagged for
RFI removal have an incorrect weighting value for those
channels.  Because data in the flagged channels are removed but not down-weighted, they
still contribute some noise to the average spectrum.  However, the number of flagged channels for RFI removal in
each scan is small enough that it does not cause any noticeable effects in
the final averaging.  The final spectrum is shown in Figure 1.        

\subsection{Arecibo Observations and Reduction}

The data for SDSS 21-00 were taken with the 305m diameter Arecibo telescope on
2006 October 27 on the L-band wide receiver and the
interim 50 MHz correlator. The correlator was configured for 9-level sampling in
2 linear polarizations with four boards. Two of the boards were centered on the 1420.4058 MHz line of 
neutral hydrogen while the other two boards were centered
on the 1667.359 and 1720.530 MHz OH lines. Radio frequency interference was too
strong at these higher frequencies to make use of the OH observations. For the
HI observations, the center frequency was set to 1338.147 MHz, the expected 
emission frequency of the galaxy. One board was set up with a 25 MHz bandwidth 
while the other was set-up with a 6.25 MHz bandwidth. In both cases there were 
2048 lags per board resulting in velocity resolutions of 2.9 and 0.7
km $s^{-1}$ respectively, before Hanning smoothing. 

Prior to the beginning of observations a test scan was taken on a blank sky
position that showed emission from the 1350/1330 MHz FAA radar. Because of this
emission, the radar blanker was used during the observations which reduces the
effective integration times by approximately a factor of 1/1.188. The
observations of the source consist of 4 on/off observations of 5 minutes each.
Each resulting on/off pair was combined, Hanning smoothed and a baseline was
fit to the result and subtracted. The eight spectra -- 4 on/off pairs with 2
polarizations each -- were then averaged and an
additional second order polynomial was subtracted from the resulting spectrum.
An additional boxcar smoothing of 25 channels or 18.1 km $s^{-1}$ was applied to the
spectrum. The final spectrum is shown in Figure 2.  

\subsection{Analysis}

Two signals were detected in the
GBT observations of Mrk~1456 (see Figure 1).  Both signals show the typical double-horned profile as expected from an
inclined disk galaxy.  One signal is located at the frequency
1.35587$\pm$0.00003 GHz, and corresponds to a redshift of
0.04759$\pm$0.00002, which matches the SDSS redshift of Mrk~1456
(0.04757$\pm$0.00008) within the errors.  The second signal, located at the frequency 1.3544$\pm$0.0001,
corresponds to a redshift of 0.04873$\pm$0.00008, which matches the
position and redshift of Mrk~1457 (SDSS redshift
0.04857$\pm$0.00009).  Mrk~1457 is located about 2.3' south of Mrk~1456.  Given the beamwidth
of 9', we note that two other members, Mrk~1458 and
SDSS~J114711.09+522653.4, located at frequencies of 1.35517 GHz (14442
km $s^{-1}$) and 1.35574 GHz (14310 km $s^{-1}$) respectively, are also within the beam.
Examination of the spectrum in Figure 1 shows that Mrk~1458 falls
in the gap between Mrk~1456 and Mrk~1457, where the HI flux
falls to zero.  SDSS~J114711.09+522653.4 falls in the middle of the signal of
Mrk~1456.  As there is no way to disentangle the two signals, there
may be some contribution of HI from SDSS~J114711.09+522653.4.
However, SDSS~J114711.09+522653.4 is an E/S0 galaxy that is 2.8' from Mrk~1456 (the beam center), so we expect the
contribution to the HI flux to be small.  The regular shape of the
spectrum also indicates that this is the case.  

The task Gmeasure in GBTIDL was
used to measure the integrated flux, S(v)dv, in $K\cdot(km s^{-1})$, the velocity width W, and systematic velocity of the
galaxy profile.  The rms noise in the data was taken as the error in
the flux. 
The results of the analysis are shown in Table 3.   

HI mass was determined from the data using \citep{ver}
\begin{equation}  
 \frac{M_{HI}}{M_{\odot}} = 2.4\times10^{5}D^2\int{S(v)dv}
\end{equation}
 where D is the luminosity distance in Mpc (see Table 2), and S(v)dv
 is the flux integrated over the line in
 $Jy\cdot(km s^{-1})$\footnote[4]{S(v)dv is converted from $K\cdot(km s^{-1})$
 to $Jy\cdot(km s^{-1})$ before $M_{HI}$ is computed}.  We include flux and
 distance errors on the HI mass measurement in Table 3. 

We estimate the dynamical mass from  
\begin{equation}
M_{dyn} = \frac{v_{rot}^2*r_{HI}}{G}
\end{equation}
where $r_{HI} = 1.5*r_{opt}$ \citep{broeils97}, $r_{opt}$ is the SDSS
r-band Petrosian radius, and $v_{rot}$
is the rotational velocity, estimated by
$v_{rot}=\frac{W_{50}}{2*sin(i)}$, where i here is the optical inclination of the galaxy.  

In addition to these basic HI properties of the galaxies, we calculate
the HI gas mass fraction, defined as $f_{gas}=M_{HI}/M_{dyn}$,  and the gas
richness, $M_{HI}/L_B$.  These
values are also given in Table 3.  All HI parameters were calculated
using the HI line width at 50$\%$ of the peak.

\section{Discussion}

\subsection{HI properties of Mrk~1456, SDSS~J211701.26-002633.7, and Mrk~1457}

We find that both Mrk~1456 and SDSS~21-00 have typical HI
properties, ie. HI mass, gas richness, gas mass fraction, for galaxies
with their given morphological types \citep{rob,sal,broeils97}.
Mrk~1457, when compared to other Seyferts, lies at the high end of the
normal range of HI masses ([0.2-9.0]x10$^9M_{\odot}$) and appears to
  have an unusually large gas-mass fraction relative to the average
  $f_{gas}$ of 0.009$\pm$0.002 for Seyferts \citep{haan}.  

Using the value of $M_{HI}^{\ast}=6.3\times10^9$ from \citet{zwaan},
Mrk~1456 and Mrk~1457 are slightly sub-$M_{HI}^*$ galaxies, while
SDSS~21-00 is twice an $M_{HI}^*$ galaxy.  Mrk~1456 and SDSS~21-00 are representative of average Sb-Sc spiral galaxies
with a fair amount of HI gas remaining, while having typcial SFR rates
of most spirals, indicative of an on-going SFR process and conversion
of their neutral gas reservoir into stars.       

\subsection{Comparison with HI-detected DLAs}

In the following, we discuss HI observations of known/candidate DLAs
and sub-DLAs, namely, SBS~1543+593, NGC~4203, and the absorbers
towards OI~363, PKS~0439-433, and PG~1216+069.  The data are collected
in Table 1.  All values listed have been recalculated using the cosmology
adopted in this paper, and all magnitudes listed are Johnson-Morgan-Cousins B-band magnitudes converted to the AB
zeropoint system, except the magnitude of the OI~363 absorber which is
a K-band magnitude.  

Presently, only one bonafide DLA has been successfully detected in HI 21~cm
emission \citep{bowen01}.  SBS~1543+593, a z=0.009 dwarf spiral \citep{schulte04} is a
quasar absorption line galaxy, first found through study of its
emission lines \citep{martel94,reimers}.  SBS~1543+593 is a low
surface brightness system that has been extensively studied in
emission and absorption \citep{rosen06,bowen01a,bowen05,schulte05}.
It has HI properties consistent with other dwarf
spirals \citep{rob,broeils97,deblok} including a low HI mass, a large
$M_{HI}/L_B$, and a small gas mass fraction. 

NGC~4203 was mapped in 21~cm \citep{van88,burstein81} and exhibits a
typical HI mass and $f_{gas}$ for the average S0, but an atypically large
gas richness \citep{rob}.  The absorbers
toward the QSOs OI~363 and PKS~0439-433 were undetected in 21~cm
emission \citep{lane00,kane01} but have 3$\sigma$ upper limits to the
HI mass.  For the OI~363 absorber, it was not optically detected with
the exception of two regions of patchy structure neat the QSO OI~363,
referred to as the ``jet'' and the ``arm'' \citep{turn01}.  Although
no optical counterpart has been found in conjunction with the
z=0.0063 absorber towards PG~1216+069, a weak 21~cm emission signal
was found at 3$\sigma$ significance within 30'' (3.8 kpc) of the QSO \citep{briggs06}.    

Figure~3 provides a comparison of several of the galaxy properties listed
in Tables 1 and 2.  For the morphological type, we assign a value to each
type ranging from 0-E to 6-Irr, with intermediates taking half-integer
values, ie. 2.5-Sab.  For the z=0.101 galaxy associated with
PKS~0439-433, \citet{kane01} report two upper limits to the HI mass,
listed in Table 1 and in Figure~3, we plot the larger value
for the upper limit to the HI mass and $M_{HI}/L_B$.  The absorbers
associated with OI~363 and PG~1216+069 are not included as no host
galaxy has, as of yet, been clearly identified with the absorbing
system.  Five objects do not carry statistical weight, but are used to just highlight some broad
trends and consistencies.  

The morphologies of the DLAs/candidates in Fig.~3 consist of one dwarf, one S0 galaxy, and three
giant late-type spirals.  This diversity in galaxy types naturally
accounts for some spread in the derived HI and optical properties shown in
Fig.~3.  With regards to their optical properties, SDSS~21-00 and
Mrk~1456 fall right in the middle of the distribution of
$L^*$, and $M_{HI}/L_B^*$ values in the given sample.  They do, however, exhibit one noteworthy property: they
have the highest HI masses of the sample.  Our observations of Mrk~1456 and SDSS~21-00 have added two candidate
DLA galaxies with morphologies and luminosities of giant late-type
spirals.  Thus they are most directly comparable to the galaxy toward
PKS~0439-433, also a late-type giant spiral.  The PKS galaxy is an Sab
class sub-DLA with an HI column just below the DLA limit, and an
impact parameter that is just outside its stellar disk.  Given that HI
column densities are higher at lower radii and in later morphological subtypes,
we expect that future observations on the QSO sightlines of Mrk~1456
and SDSS~21-00 might reveal HI column densities larger than the DLA limit, as both are
later subtypes than the PKS galaxy and both have sightlines to their
QSOs that lie inside their stellar disks.  The properties of Mrk~1456 and SDSS~21-00 agree within
the distributions of expected optical properties of z=0 DLAs. 
                         
\citet{rosen06} found that SBS 1543+593 is in a small galaxy
group with three other companions, a disk galaxy and
two dwarfs, all of which are of low luminosity.  All three companions
are gas rich systems, with low HI mass ($<10^9M_{\odot}$),
and average gas content for disk and late-type dwarf galaxies
\citep[see][Table 2]{rosen06}.  This group, and other small groups like it, are
likely to be fairly common and important in the contribution to
the DLA population.  Furthermore, if young, gas-rich galaxy groups are
more prevalent at higher redshifts, then these systems might be
important to the DLA cross-section.  From a search using the NASA/IPAC
extragalactic database (NED), none of the other DLAs that have been
observed in 21~cm have associated groups, but faint dwarfs could have
been missed.

\citet{merch} completed a sample of galaxy groups in the DR3 of
the Sloan Digital Sky Survey (SDSS).  Mrk~1456 is in one such group,
with four other members.  Mrk~1457 is one such member.  All
other galaxies in the group are contained within the beam of the GBT,
centered on Mrk~1456, but Mrk~1457 was the only other member to be detected in HI
21~cm emission.  \citet{merch} also show that SDSS~21-00 is in a group, consisting of seven
other members.  Table 4 lists a few global properties of the group,
ie. the group coordinates, group systematic velocity, group velocity width, virial mass and
radius of the group, as calculated by \citet{merch} and Table 5 lists
each group's members and their properties.  As both galaxies are
situated in groups, the question of misidentification of the DLA host
galaxy becomes an issue.  Traditionally, identifying DLA hosts
has proven difficult since either no obvious galaxy is visible or
identifying the correct host in a group is difficult as the galaxy giving rise to the line may not be
the brightest or closest galaxy to the QSO. This is not an issue with
Mrk~1456 or SDSS~21-00.  For both systems, the QSO lies within the
stellar disk of the galaxy, and in both groups, the closest
companion outside the named galaxy here lies more than 120 kpc from
the QSO.    

Without HI observations of the galaxies in these groups, it is
hard to determine the richness of the group and
the environment these galaxies live in.  We use u-r color and
morphological type as a rough indicator of the gas content of the
other galaxies in these groups. The reliability, however, of using
optical properties as a predictor of gas content is uncertain as
HI-selected galaxies differ more in their optical
properties than optically-selected ones, the exception being the most
blue galaxies, as they have retained most of their primordial gas.
Figure 5 shows how galaxies of different morphologies segregate in
color and concentration.  We plot the galaxies from \citet{park05}
overlaid with the Mrk~1456 group (black circles) and the SDSS~21-00
group (black diamonds).  The galaxies in the Mrk~1456 group, with the
exception of Mrk~1456, fall in the region of the diagram occupied by
early types.  The galaxies in the SDSS~21-00 galaxy group, on the other hand, are
predominantly Sb-Sc morphological types with only two E/S0 galaxies.
\citet{kannap04} showed that there is a relationship between u-r color
and gas richness.  Galaxies with u-r $<$ 1.5 are gas-rich, 1.5 $<$ u-r $<$ 2.5 are intermediate in
gas richness, and those with u-r $>$ 2.5 are generally gas-poor. The u-r
colors of the galaxies in the two galaxy groups are listed in Table 5.
These results indicate that both the Mrk~1456 and SDSS~21-00 galaxy
groups are intermediate in their gas-richness.  

The total HI mass of all 4 galaxies
detected in the SBS~1543+593 galaxy group adds up to
$<$3$\times10^{9}M_{\odot}$.  With two of four galaxies detected in the
Mrk~1456 group, its HI content is already much larger,
$\sim$12$\times10^{9}M_{\odot}$.  The single galaxy detected in the
SDSS~21-00 galaxy group has an HI mass of
$\approx$14$\times10^{9}M_{\odot}$, much larger than the entire
SBS~1543+593 galaxy group.  

\subsection{Comparison with HI-detected galaxies}

\citet{west05} used SDSS to identify optical counterparts of
HI-selected galaxies from HIPASS and found that HI-selection yields a high fraction of late-type
galaxies.  For HI-selected spirals, \citet{west08} give the following
median properties: $M_{HI}/L_B$ = 0.5, $M_{HI}$ = $M_{HI}^*$,
and $L_B$ = 0.6 $L_B^*$.  Our Ca~II
absorbing galaxies appear to have properties that are comparable to
those of HI-selected galaxies.  

\citet{rosen03} and \citet{ryan03} also looked at HI-selected galaxies, and
focused more specifically on the question of how these contribute to
the DLA cross-section.  \citet{rosen03} find a tight correlation
between the expected DLA cross-section and the HI mass.  Their
equation 3 predicts the following DLA cross-sections for Mrk~1456 and
SDSS~21-00: 853.7 $kpc^2$ and 2058.4 $kpc^2$.  Using these sizes to
estimate the average column density for the galaxies, we find for Mrk~1456,
$N_{HI}=8.23\times10^{20}cm^{-2}$, and for SDSS~21-00,
$N_{HI}=8.25\times10^{20}cm^{-2}$.   Rosenberg $\&$ Salzer (in prep.) looked at the
predicted optical properties for z=0 DLAs from an HI-selected sample
and find that the most common morphological type is spiral (45$\%$) and that
50$\%$ of the DLAs should have an $M_{HI}/L_B$ $>$ 0.77 (2 of the 6 galaxies
from the Rosenberg $\&$ Salzer (in prep.) sample are from this sample).
\citet{ryan03} also find that gas-rich, late-type spirals contribute most to the DLA
cross-section.    

Figure 4 shows the HI mass contribution to
dN/dz, the number of expected systems per unit redshift, for z$\sim$0,
for a sample of HI-selected galaxies with an HI column larger than
the DLA limit from the Arecibo Dual-Beam Survey (ADBS, see
\citet{rosen03} for details).  The figure shows that 50$\%$ of the cross-sectional area is from galaxies
with HI masses between 2.9$\times10^8$ and 3.5$\times10^9$.  The
red-hatched histogram overlayed on top is the HI mass distribution of
the objects listed in Tables 1 and 2, with the exception of the OI~363
absorber.  The right hand axis lists the values for this histogram,
N, the number of galaxies.  One might expect the HI mass distribution
of DLAs in general to follow the same HI mass distribution of
HI-selected galaxies expected to be DLAs.  It should be noted
that the two galaxy/QSO pairs we add here were not selected using the
traditional selection method for quasar absorption line systems,
ie. through UV absorption lines.  Thus, they introduce some bias in
the HI mass distribution as they are giant spirals optically-selected
from SDSS with quasars projected within their stellar disks.  However,
even with this caveat and low statistics, it appears the HI mass
distribution of the (candidate) DLAs mimics the distribution of
HI-selected galaxies expected to be DLAs.     

\citet{zwaan} looked at an
optically-selected sample of the local galaxy population (z$\sim$0),
mapped in 21~cm emission, in order to connect them to the low-redshift
(z$<$1) DLA population and calculated expected probability
distribution functions of different properties for the low-redshift
DLAs.  They find that, for expected DLA systems, most will be
late-type, with a median impact parameter of 8.0 kpc$^{+10.8}_{-5.3}$,
median log $M_{HI}$ of 9.3$^{+0.5}_{-0.7}$, and median
$M_B$=-17.98$^{+2.7}_{-2.0}$.  Mrk~1456 and SDSS~21-00 are indeed
late-types within the range for the expected impact parameter and have
values for HI mass and absolute magnitude that lie close to the
expected values, on the high end.
               
\section{Conclusion}

We report on the HI 21~cm emission observations of two galaxies,
Mrk~1456 and SDSS~J211701.26-002633.7.  Both are average disk
galaxies and candidate DLAs, based on the strengths of their Ca~II~K
EW's and their low impact parameters to the QSO.  In terms of their HI
properties, both galaxies have typical
HI properties for their morphological types with $M_{HI}$ $\sim$
$M_{HI}^*$, average gas richness, and gas mass fractions that are
representative of disk galaxies.  

These data add to the total number of strong candidate DLAs with HI observations, in the
local Universe.  Compared with other HI-detected DLAs, Mrk~1456 and SDSS~21-00 are
similar in most respects, with comparatively higher HI masses. 
When compared to the expected properties of z=0 DLAs from HI-detected
galaxies, we find that Mrk~1456 and SDSS~21-00 fall within range of the
expected HI mass distribution, median impact parameter, and absolute
magnitudes for local DLAs.  We also find that both
Mrk~1456 and SDSS~21-00 reside in groups, adding two DLAs in gas-rich
groups to the known population.  This may indicate a common phenomenon
with regards to the environment of DLAs. 

In spite of many years of observational effort, associated galaxies have only been
identified in a very small fraction of known DLA systems. Therefore, the nature of
the evolving DLA galaxy population has continued to be under investigation, both
observationally and theoretically. Historically, two very different scenarios were
developed to explain what kind of galaxies produce DLAs. In the explanation put
forth by \citet{wolf2}, DLAs are associated with massive spiral galaxies. In
the model proposed by \cite{haen98}, DLAs instead arise in low-mass dwarf
galaxies. 

The numerical and semi-analytical galaxy formation simulations presented over the
last decade have modified our view of DLAs.  Current models suggest that the median properties of 
DLA galaxies and their range evolve as a function of cosmic time.  In this scenario, dwarfs make up most of the DLA cross-section at high redshifts, with massive disk galaxies contributing more as the Universe evolves.  \cite{johan06} find that higher mass systems make an increasing contribution to the DLA population as redshift decreases. Specifically, \cite{johan06} predict a median DLA 
Hydrogen mass of $M_{HI}$ = 2$\times$ $10^9M_{\odot}$ at z=0; in good agreement
with the results for HI-selected galaxies \citep{zwaan}, and what we find here
(cf. Fig.~4).  Thus, although Mrk 1456 and SDSS~21-00 are
case studies of bright QSOs seen through the disks of nearby spiral galaxies, we do
not conclude that they support the original \citet{wolf2} picture.  Rather, we interpret our results to be compatible with the current theoretical models,  as galaxies contributing to the fraction of disk galaxies that make up the DLA cross-section at low redshift.  We conclude that our result further strengthens the picture that the
local galaxy population, in its variety of gas rich galaxy types that randomly
intercept QSO sightlines, can explain the properties of the low-redshift DLAs.  

\acknowledgments 
\indent We are grateful to Toney Minter, Karen O'Neil, and Jay Lockman at the
GBT for their help in the observations and reduction of the data.  The
National Radio Astronomy Observatory is a facility of the National
Science Foundation operated under cooperative agreement by Associated
Universities, Inc. \\ 
\indent We thank Chris Salter and Tapasi Ghosh for handling the Arecibo
observations.  The Arecibo Observatory is part of the National
Astronomy and Ionosphere Center, which is operated by Cornell
University under a cooperative agreement with the National Science
Foundation.  \\ \indent We wish to thank Andrew West for making details of his Ph.D thesis
available to us. \\
\indent We also made use of data obtained at the Gemini Observatory, which is operated by the
Association of Universities for Research in Astronomy, Inc., under a cooperative agreement
with the NSF on behalf of the Gemini partnership: the National Science Foundation (United
States), the Science and Technology Facilities Council (United Kingdom), the
National Research Council (Canada), CONICYT (Chile), the Australian Research Council
(Australia), Ministério da Ciência e Tecnologia (Brazil) and SECYT
(Argentina). \\
\indent  Funding for the SDSS and SDSS-II has been provided by the Alfred
P. Sloan Foundation, the Participating Institutions, the National
Science Foundation, the U.S. Department of Energy, the National
Aeronautics and Space Administration, the Japanese Monbukagakusho, the
Max Planck Society, and the Higher Education Funding Council for
England.  The SDSS Web Site is http://www.sdss.org/.  The SDSS is managed by the
Astrophysical Research Consortium for the Participating
Institutions. The Participating Institutions are the American Museum
of Natural History, Astrophysical Institute Potsdam, University of
Basel, University of Cambridge, Case Western Reserve University,
University of Chicago, Drexel University, Fermilab, the Institute for
Advanced Study, the Japan Participation Group, Johns Hopkins
University, the Joint Institute for Nuclear Astrophysics, the Kavli
Institute for Particle Astrophysics and Cosmology, the Korean
Scientist Group, the Chinese Academy of Sciences (LAMOST), Los Alamos
National Laboratory, the Max-Planck-Institute for Astronomy (MPIA),
the Max-Planck-Institute for Astrophysics (MPA), New Mexico State
University, Ohio State University, University of Pittsburgh,
University of Portsmouth, Princeton University, the United States
Naval Observatory, and the University of Washington. \\ 
\indent This research has made use of NASA's Astrophysics Data System
and of the NASA/IPAC Extragalactic Database (NED) which is operated by
the Jet Propulsion Laboratory, California Institute of Technology,
under contract with the National Aeronautics and Space Administration.


\clearpage
\onecolumn
\begin{figure}
\includegraphics[angle=90,scale=0.7]{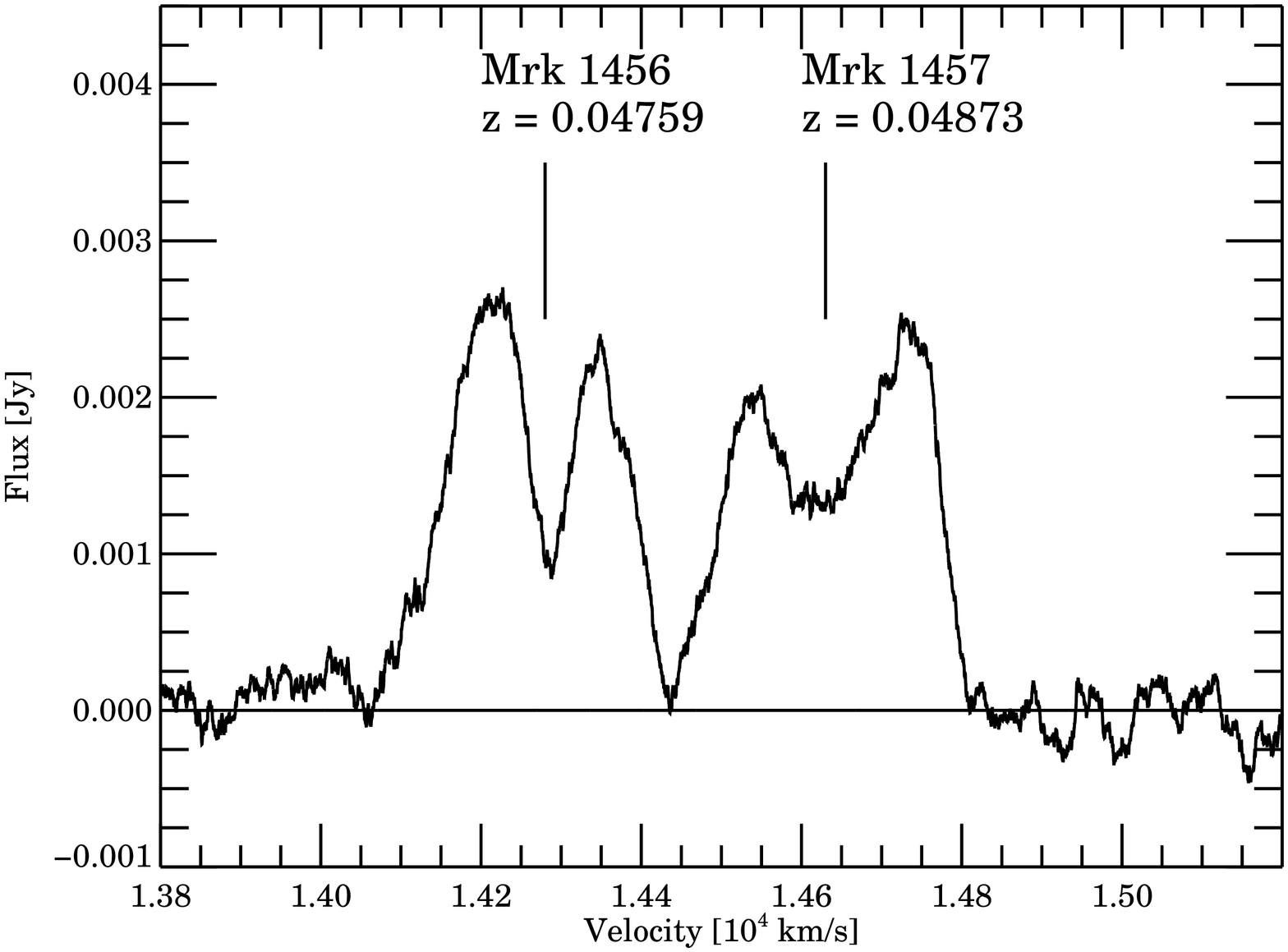}
\caption{HI 21~cm spectrum from GBT showing Mrk 1456 and Mrk 1457
  \label{fig1}}
\end{figure}

\clearpage
\begin{figure}
\includegraphics[angle=90,scale=0.7]{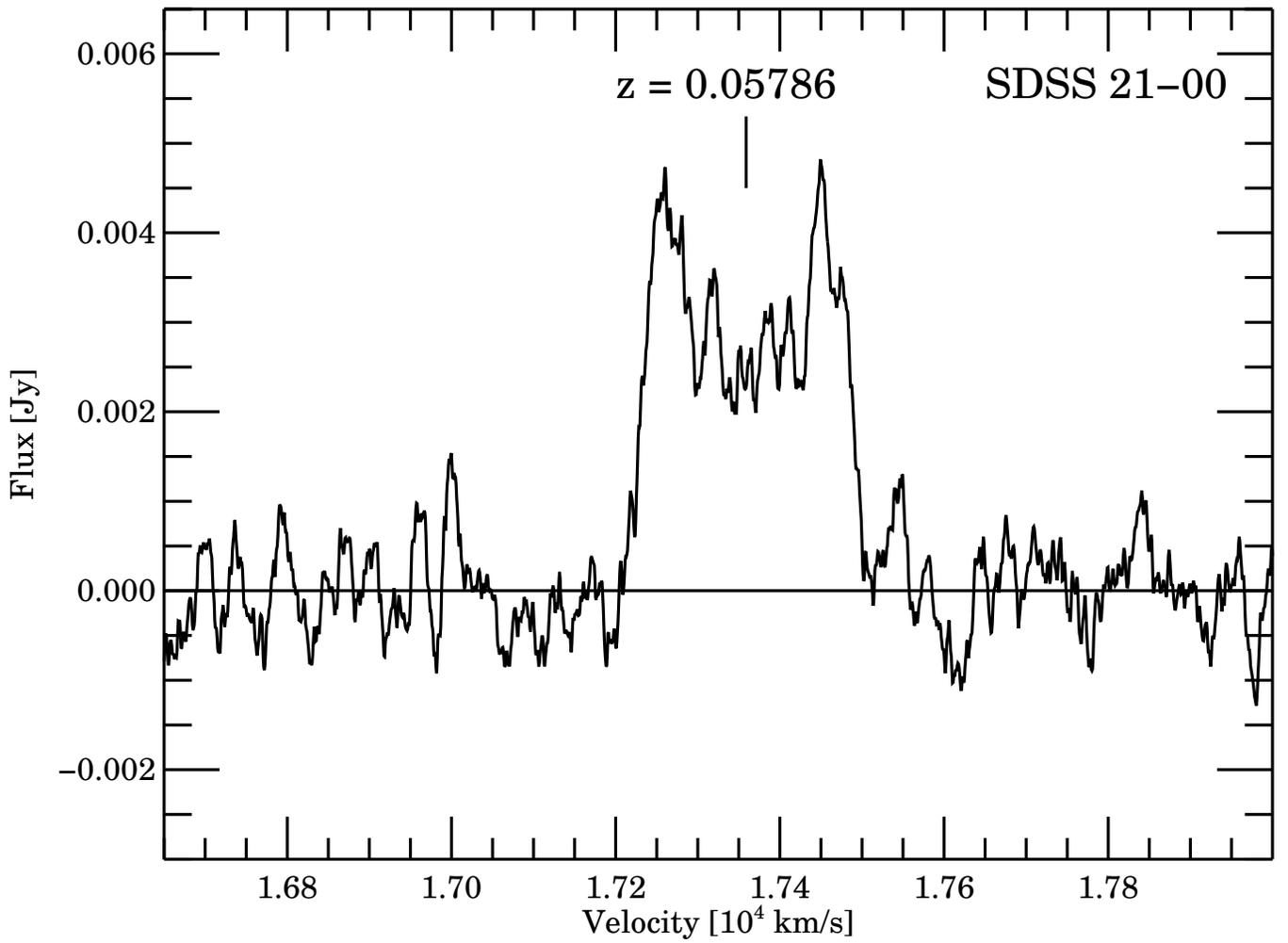}
\caption{HI 21~cm spectrum from Arecibo showing SDSS~J211701.26-002633.7}
\end{figure}

\clearpage
\begin{figure}
\includegraphics[angle=0, scale=0.8]{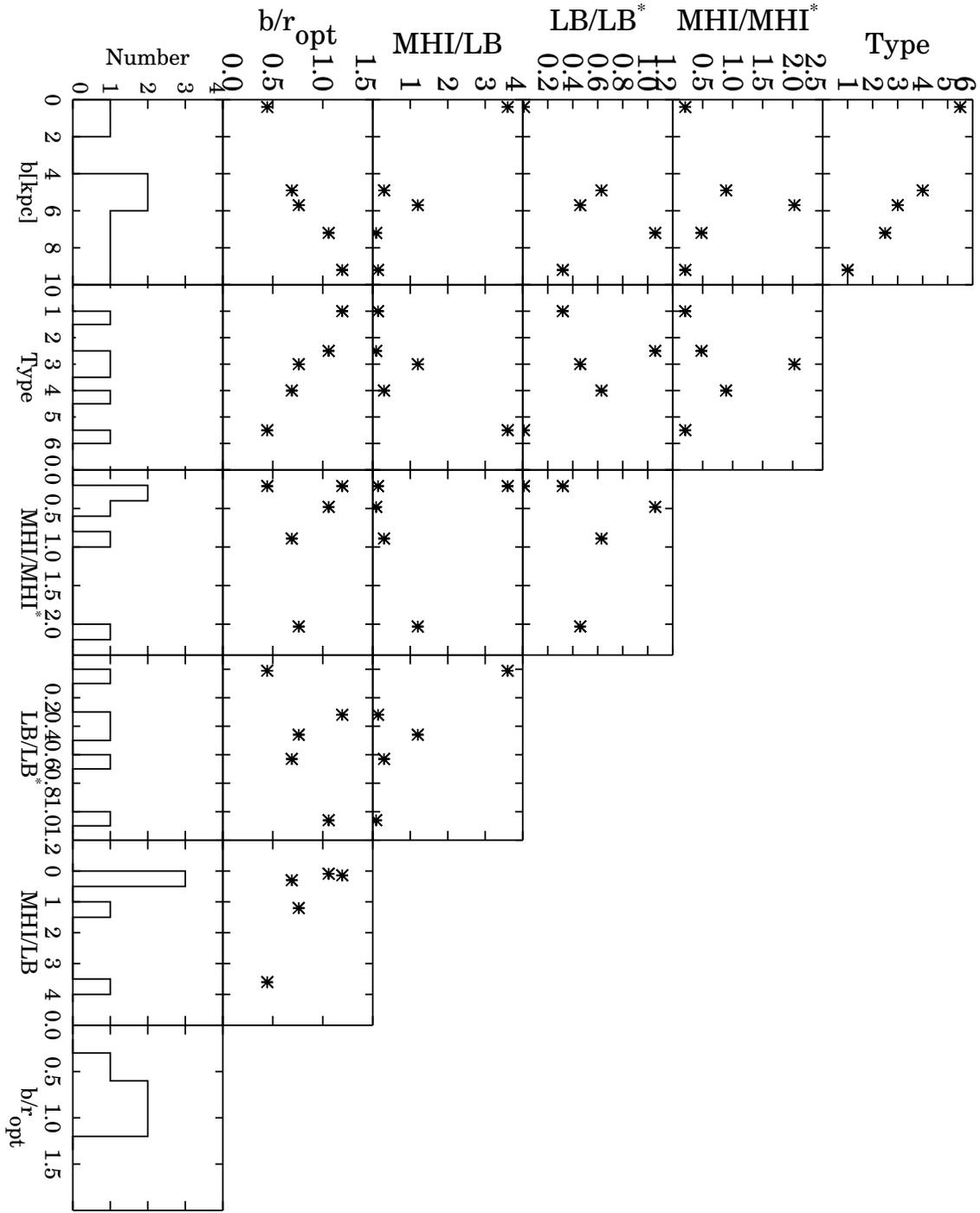}
\caption{Histograms and plots of impact parameter, morphological type, $M_{HI}$, $L_B$,
  $M_{HI}/L_B$, and ratio of impact parameter to optical radius of the HI DLA
  sample.  The morphological type is represented with numbers from 0-6
  for types E-Irr, respectively. $M_{HI}$ is in units of $M_{HI}^*$
  and $L_B$ is in units of $L_B^*$.  The absorbers associated with
  OI~363 and PG~1216+069 are omitted from the diagram.}
\end{figure}

\clearpage
\begin{figure}
\includegraphics[angle=0,scale=0.9]{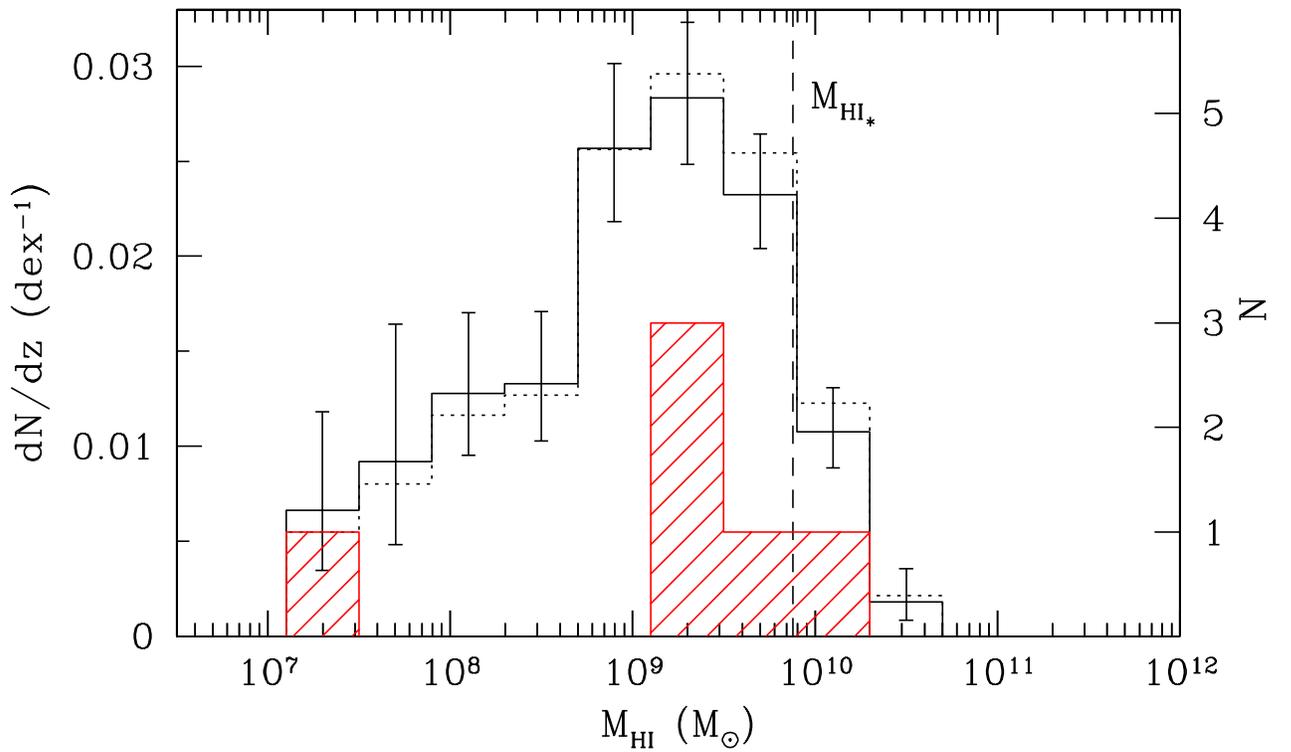}
\caption{dN/dz as a function of HI
  mass.  The open solid and dashed histograms show the distribution for a HI-selected sample of galaxies, representative of z=0
  DLAs, from \citet{rosen03}.  The red hatched histrogram shows the HI masses for the DLA
  systems listed in Tables 1 and 3.  The right-hand axis shows the number of
  galaxies.  The dashed line indicates $M_{HI}^*$.}
\end{figure}

\clearpage
\begin{figure}
\includegraphics[angle=0, scale=2.0]{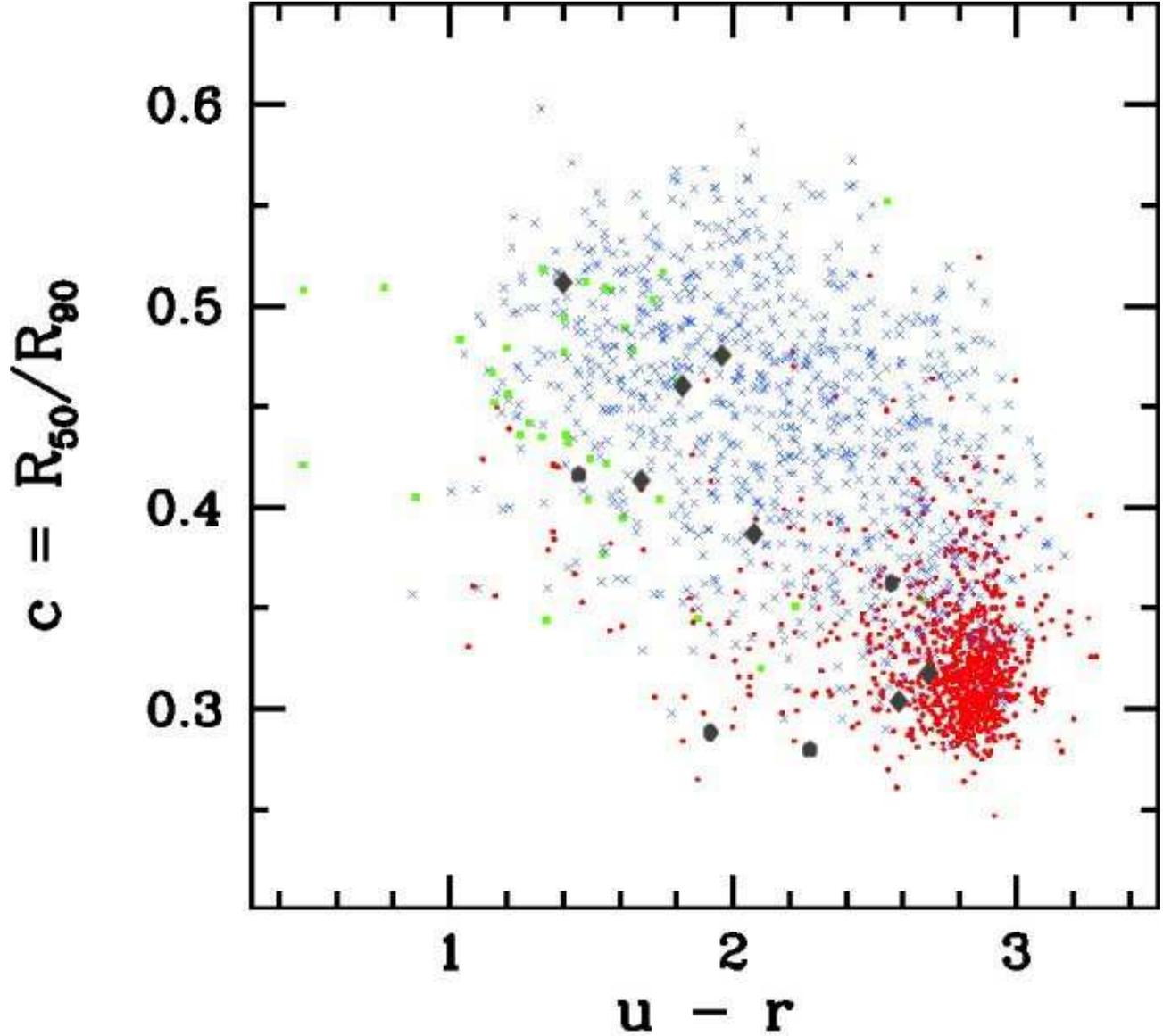}
\caption{Galaxy concentration index versus optical color.  The red
  circles show early type galaxies, green squares show intermediate
  types, and blue circles show late types from \citet{park05}.  The black dots are the
  Mrk~1456 galaxy group members and the black diamonds are the
  SDSS~21-00 group memebers.  Most members from the Mrk~1456 group
  tend to be early-types, while most from the SDSS~21-00 group
  tend to be late-types, suggesting the latter group is overall more
  gas-rich than the former.}  
\end{figure}




\clearpage
\begin{table}[tbp]
\renewcommand{\arraystretch}{1.5}
{\footnotesize
\caption{Optical and HI properties of Known/Candidate DLA Galaxies}
\begin{tabular}{ccccccc}
\tableline\tableline
Property & SBS~1543+593 & NGC~4203 & 'Galaxy' & ``Jet''$^a$ & ``Arm'' & -- \\
\tableline
type$^b$ & Sdm & SO & Sab & LSB & LSB & --\\
z & 0.009$\pm$0.001 & 0.0036$\pm$0.00001 & 0.101$\pm$0.001 &
0.09118$\pm$0.00001 & 0.09118$\pm$0.00001 & 0.00632$\pm$0.00002\\
$D_L$[Mpc] & 38.3$\pm$4.3 & 15.25$\pm$0.4 & 460.0$\pm$4.6 &
412.46$\pm$0.04 & 412.46$\pm$0.04 & 26.84$\pm$0.08\\
QSO & HS~1543+5921 & Ton~1408 & PKS~0439-433 & OI~363 & OI~363 & PG~1216+069\\
$z_{QSO}$ & 0.807$\pm$0.002 & 0.6156$\pm$0.0002 & 0.593$\pm$0.003 &
0.631$\pm$0.002 & 0.631$\pm$0.002 & 0.3313$\pm$0.0003\\
b[''] & 2.4 & 126 & 3.9 & $<$2 & $<$2 & $<$30\\
b[kpc] & 0.4 & 9.2 & 7.2 & $<$3.4 & $<$3.4 & $<$3.9 \\
r['']$^c$ & 4.8 & 105 & 3.7 & 4 & 4 & -- \\
r[kpc]$^c$ & 0.9 & 7.7 & 6.8 & 6.7 & 6.7 & -- \\
axis ratio b/a & 0.777 & 0.880 & 0.556 & -- & -- & -- \\
$i^d$[$^{\circ}$] & 40 & 29 & 58 & -- & -- & -- \\
$M_B^{k,f,i: e}$ & -16.1$\pm$0.2 & -19.65$\pm$0.01 & -20.96$\pm$0.08 & -14.72$^f$ & -15.12$^f$ & -- \\
$L_B[10^{10}L_{\odot}]$ & 0.04$\pm$0.01 & 0.96$\pm$0.02 & 3.2$\pm$0.2& 0.01 & 0.02 & -- \\
$L_B^*$ & 0.01 & 0.32 & 1.06$\pm$0.07 & 0.0001 & 0.0002 & -- \\
$SFR[M_{\odot}yr^{-1}]^g$ & 0.006 & 0.16 & -- & -- & -- & --\\
$W_{50}[km s^{-1}]$ & 75 & 243 & -- & -- & -- & -- \\
$M_{HI}[10^9M_{\odot}]$ & 1.3$\pm$0.1 & 1.3 & $<$(2.15-3.02) & $<$3 &$<$3 & (0.006-0.017) \\
$M_{HI}^*$ & 0.21 & 0.21 & $<$0.34-0.48 & $<$0.48 & $<$0.48 & (0.001-0.002) \\
$M_{dyn}[10^{11}M_{\odot}]$ & 0.07 & 1.2 & -- & -- & -- & -- \\
$\frac{M_{HI}}{L_B}$ & 3.6$\pm$0.9 & 0.14$\pm$0.01 & $<$(0.06-0.09) & $<$30 & $<$20 & -- \\
$\frac{M_{HI}}{M_{dyn}}$ & 0.19 & 0.01 & -- & -- & -- & -- \\
$log N_{HI}[cm^{-2}]^h$ & 20.42$\pm$0.04 & 20.34 & 19.85$\pm$0.10 &
21.18$\pm$0.06 & 21.18$\pm$0.06 & 19.32$\pm$0.03 \\
Refs$^i$ & 1,2,3,4 & 5,6 & 7,8,9,10 & 11,12,13 & 11,12,13 & 14,15,16\\
\tableline
\end{tabular}
\tablenotemark{a}{Cherinka,K\"{o}nig,$\&$ Schulte-Ladbeck attempted detection of H$\alpha$ of the ``jet'' with the GMOS-N IFU , but none was found, 2005, unpublished}\newline
\tablenotemark{b}{morphological classification}\newline
\tablenotemark{c}{radius estimates for: SBS~1543 $\&$ NGC~4203 - from diameter of 25th B-band isophote; PKS~0439-433 galaxy - radius towards QSO,measured from V band image; OI~363 - using K band estimate 8''}\newline
\tablenotemark{d}{SBS~1543 - HI incl. measured from rotation curve; NGC~4203 - HI inclination of inner disk; PKS~0439-433 galaxy - optical incl. from K-band}\newline
\tablenotemark{e}{absolute B-band magnitude: superscripts k, f, and i denote the magnitude is k-corrected, foreground extinction corrected, and internal extincton corrected via \citet{tully98}, respectively \newline}
\tablenotemark{f}{K-band magnitudes, not internal extinction
  corrected, converted to luminosity using \citet{love} solar value\newline}
\tablenotemark{g}{SBS~1543 SFR - total $H_{\alpha}$ star-formation rate from \cite{schulte04}; NGC~4203 SFR - FIR star-formation rate calculated with 60 and 100 $\mu$m flux densities available from the Infrared Astronomical Satellite (IRAS) along with equations from \citet{helou88} and \citet{hopkins} \newline}
\tablenotemark{h}{$N_{HI}$ source: SBS~1543 - Ly-$\alpha$ line; NGC~4203 - X-ray absorption; PKS~0439-433 galaxy - Ly-$\alpha$ line; Jet and Arm near OI~363 - Ly-$\alpha$ line; Unknown galaxy with PG~1216+069 - HI~21cm emission\newline}
\tablenotemark{i}{1-\cite{bowen01}; 2-\cite{rosen06}; 3-\cite{schulte04}; 4-\cite{cheng};\\
    5-\cite{miller99}; 6-\cite{van88}; 7-\cite{kane01}; 8-\cite{petit96}; \\
    9-\cite{chen03}; 10-\cite{chen05}; 11-\cite{lane00}; 12-\cite{turn01}; 13-\cite{rao98};\\
    14-\cite{briggs06}; 15-\cite{tripp05}; 16-\cite{kane05}}
}
\end{table}


\clearpage
\begin{table}[tbp]
\renewcommand{\arraystretch}{1.5}
{\small
\caption{Optical Properties of Mrk~1456 and SDSS~21-00}
\begin{tabular}{ccc}
\tableline\tableline
Property & Mrk~1456 & SDSS~21-00 \\
\tableline
Type & Sc & Sb \\
z & 0.04757$\pm$0.00008 & 0.05792$\pm$0.00009 \\
$D_L$[Mpc] & 208.5$\pm$0.3 & 255.8$\pm$0.4 \\
QSO & SDSS~J114719.89+522923.1 & SDSS~J211701.31-0026.38.8 \\
$z_{QSO}$ & 1.99 & 1.14 \\
b[''] & 5.3 & 5.1 \\
b[kpc] & 4.9 & 5.7 \\
r[''] & 7.7 & 6.7\\
r[kpc] & 7.1 & 7.5 \\ 
axis ratio b/a & 0.724 & 0.638\\
i[$^{\circ}$] & 45 & 52 \\
$M_r^{k,f}$ & -20.58$\pm$0.01& -20.14$\pm$0.01\\
$M_u^{k,f}$ & -19.14$\pm$0.02& -18.55$\pm$0.04\\
$SFR_u[M_{\odot}yr^{-1}]$ & 3.1$\pm$0.2 & 5.6$\pm$1.2\\
$M_B^{k,f,i}$ & -20.40$\pm$0.09$^a$& -20.07$\pm$0.09\\
$L_B[10^{10}L_{\odot}]^b$ & 1.92$\pm$0.16& 1.41$\pm$0.12\\
$L_B^*$ & 0.63$\pm$0.05 & 0.46$\pm$0.04 \\
REW(CaII) & 1.24$\pm$0.15 & 1.1$\pm$0.2 \\
H$\alpha$ [$10^{-16}$ergs $cm^{-2}$ $s^{-1}$]& 65.65$\pm$0.3 & 20.3$\pm$0.2\\
$L_{H\alpha}$ [$10^{40}$ergs $s^{-1}$]$^c$ & 8.64$\pm$1.69 & 13.8$\pm$0.2\\
$N_{HI}d$ [$10^{22}$ $cm^{-2}$]$^d$ & 1.8 & 3.0\\
H$\beta$ [$10^{-16}$ergs $cm^{-2}$ $s^{-1}$]& 15.15$\pm$0.2 & 2.8$\pm$0.2\\
$[NII]\lambda$6550 [$10^{-16}$ergs $cm^{-2}$ $s^{-1}$]& 6.8$\pm$0.2 & 2.09$\pm$0.14\\
$[NII]\lambda$6585 [$10^{-16}$ergs $cm^{-2}$ $s^{-1}$]& 19.40$\pm$0.3 & 5.29$\pm$0.15\\
$[OII]\lambda$3727 [$10^{-16}$ergs $cm^{-2}$ $s^{-1}$]& 35.25$\pm$0.2 & 9.3$\pm$0.8\\
$[OIII]\lambda$5008 [$10^{-16}$ergs $cm^{-2}$ $s^{-1}$]& 9.33$\pm$0.1 & 2.1$\pm$0.2\\
12+log(O/H)[O3N2]$^e$ & 8.6$\pm$0.1$\pm$0.14 & 8.58$\pm$0.02$\pm$0.14\\ 
12+log(O/H)[N2]$^e$ & 8.5$\pm$0.2$\pm$0.18 & 8.56$\pm$0.01$\pm$0.18\\
12+log(O/H)[R23]$^e$ & 8.7$\pm$0.2$\pm$0.10 & 8.10$\pm$0.11$\pm$0.10 \\
\tableline
\end{tabular}
\tablenotetext{a}{this magnitude differs from the
  value (-21.0$\pm$0.02) of \citet{koenig} in that we use Poggianti's
  kcorrection code and an internal extinction correction based on
  inclination and HI velocity width, while they use Blanton's
  kcorrection code (v3.2) \citet{blant} and correct for internal extinction using the
  Balmer-line ratio and the method of \citet{cardelli}}\newline
\tablenotetext{b}{using colors from \citet{holmberg06} and converting to AB zeropoint with \citet{frei95}}
\tablenotetext{c}{dereddened luminosity}
\tablenotetext{d}{nuclear HI column density derived from $SFR_{H\alpha}$}
\tablenotetext{e}{the two errors listed are our measured error and the systematic error in the calibration for each of the three techniques used, respectively}
}
\end{table}


\clearpage
\begin{landscape}
\begin{table}[tbp]
\renewcommand{\arraystretch}{1.5}
{\small
\caption{HI properties of Mrk~1456, SDSS~21-00, and Mrk~1457}
\begin{tabular}{ccccccccccc}
\tableline
Galaxy & Sdv & $W_{50}$ & $W_{20}$ & $v_{sys}$ & $M_{HI}$ & $M_{HI}^*$ &
$M_{dyn}$ & $\frac{M_{HI}}{L_B}$ & $\frac{M_{dyn}}{L_B}$ & $\frac{M_{HI}}{M_{dyn}}$\\
     & [$K\cdot(km\,s^{-1}$]& [$km\,s^{-1}$]&[$km\,s^{-1}$]&
[$km\,s^{-1}$]& [$10^9M_{\odot}$]& & [$10^{11}M_{\odot}$] & & & \\
\tableline
Mrk~1456 & 1.0808$\pm$0.0016 & 271.3 & 322.53 & 14262.1 &
5.64$\pm$0.002 & 0.89 & 0.9$\pm$0.4 & 0.30$\pm$0.02 & 4.8$\pm$0.6 & 0.062$\pm$0.006\\
SDSS~21-00 & -- & 255.6 & 281.8 & 17357.7 & 13.6$\pm$0.3 & 2.03 &
0.69$\pm$0.15 & 1.2$\pm$0.1 & 6.1$\pm$0.9 & 0.20$\pm$0.03 \\
Mrk~1457 & 1.0953$\pm$0.0012 & 294.1 & 344.94 & 14631.1 &
5.97$\pm$0.02 & 0.94 & 2.1$\pm$0.7 & 0.14$\pm$0.01 & 5.0$\pm$0.4 & 0.029$\pm$0.002\\
\tableline
\end{tabular}
\tablenotetext{1}{measurements made using Gmeasure mode 2, which takes
  20$\%$ or 50$\%$ of the highest peak, and all
  properties calculated using the $W_{50}$ velocity width}
}
\end{table}
\end{landscape}

\clearpage
\newpage
\begin{table}[tbp]
\renewcommand{\arraystretch}{1.5}
{\small
\begin{center}
\caption{Properties of the Mrk 1456 \& 'SDSS 21-00' Galaxy Group}
\begin{tabular}{ccccccccc}
\tableline
Group & COM & COM $v_{rad}$ & W & $M_{vir}$ & No. of Members & $R_{vir}$ &&\\ 
 & [hh:mm:ss, dd:mm:ss] & [km $s^{-1}$] & [km $s^{-1}$] & [$10^{13} M_{\odot}$] & & [Mpc] &&\\  
\tableline
Mrk 1456 & 11:47:23.76, +52:27:32.40 & 14390.04 & 172.275 &
0.7352 & 4 & 0.3551 &\\
SDSS J211701.26-002633.7 & 21:17:18.96, -00:21:14.40 & 17339.520 & 70.697 &
0.3911 & 7 & 1.1219 &\\
\tableline
\end{tabular}
\end{center}
\tablenotetext{1}{table lists overall group properties from
  \citep{merch}: \newline columns are group coordinates, group systematic
  velocity, group velocity width, virial mass, $\#$ of members, and
  virial radius}
}
\end{table}

\begin{table}[tbp]
\renewcommand{\arraystretch}{1.5}
{\small
\begin{center}
\caption{Properties of Groups Members}
\begin{tabular}{ccccccccc}
\tableline
\multicolumn{9}{c}{Mrk 1456 Galaxy Group} \\
\tableline
Members & Ra & Dec & $v_{sys}$ & z & b & $b_{COM}^a$ & Type & u-r\\
        & [hh:mm:ss] & [dd:mm:ss] & [km $s^{-1}$] & & ['] & ['] & &\\ 
\tableline
Mrk 1456 & 11:47:20.20 & +52:29:18.60 & 14261$\pm$24 &
0.04757$\pm$0.00008 & --- & 1.9 & Sc & 1.47\\ 
Mrk 1457 & 11:47:21.61 & +52:26:58.31 & 14562$\pm$28 &
0.04857$\pm$0.00009 & 2.3 & 0.7 & Sy2 & 1.94\\
Mrk 1458 & 11:47:41.67 & +52:26:55.88 & 14431$\pm$27 &
0.04814$\pm$0.00009 & 4.0 & 2.8 & Sa & 2.29\\
SDSS J114711.09+522653.4 & 11:47:11.10 & +52:26:53.40 & 14308$\pm$46 &
0.0477$\pm$0.0002 & 2.8 & 2.0 & E/S0 & 2.57\\
\tableline
\multicolumn{9}{c}{SDSS J211701.26-002633.7 Galaxy Group} \\
\tableline
Members & Ra & Dec & $v_{sys}$ & z & b & $b_{COM}^a$\\
        & [hh:mm::ss] & [dd:mm:ss] & [km $s^{-1}$] & & ['] & [']\\
\tableline
SDSS J211701.26-002633.7 & 21:17:01.27 & -00:26:33.80 & 17363$\pm$27 &
0.05792$\pm$0.00009 & -- & 6.9 & Sbc & 1.69\\
SDSS J211716.84-002715.6 & 21:17:16.85 & -00:27:15.60 & 17423$\pm$28 &
0.05812$\pm$0.00009 & 4.0 & 6.0 & Sbc & 1.83\\
SDSS J211715.44-002435.5 & 21:17:15.44 & -00:24:35.50 & 17310$\pm$40 &
0.0577$\pm$0.0001 & 4.1 & 3.5 & E/S0 & 2.60\\
SDSS J211709.94-003028.3 & 21:17:09.94 & -00:30:28.30 & 17365$\pm$17 &
0.05792$\pm$0.00006 & 4.5 & 9.5 & Sc & 1.41\\
SDSS J211723.72-001845.9 & 21:17:23.72 & -00:18:45.90 & 17391$\pm$16 &
0.05801$\pm$0.00005 & 9.6 & 2.7 & Sbc & 1.97\\
SDSS J211731.47-001246.9 & 21:17:31.47 & -00:12:46.90 & 17308$\pm$46 &
0.0577$\pm$0.0002 & 15.7 & 9.0 & S0 & 2.70\\
SDSS J211733.50-000805.0 & 21:17:33.50 & -00:08:05.00 & 17217$\pm$19 &
0.05743$\pm$0.00007 & 20.2 & 13.7 & Sc & 2.09\\
\tableline
\end{tabular}
\end{center}
\tablenotetext{a}{distance from Center of Mass of the group}
}
\end{table}


\end{document}